\documentstyle[12pt]{article}
\begin{document}
\textwidth  165mm
\textheight 215mm
\topmargin -1.5cm % for epson - 3.5cm  for laserjet appr. -12pt
\oddsidemargin -0.1cm
\evensidemargin -0.1cm

\centerline{Contributed paper for the}
\centerline{{\it VIII Marcel Grossmann Meeting on General Relativity,}
Jerusalem, June 23-27, 1997}
\centerline{{\it Foundations of Physics Letters}, in press}
\centerline{PACS 11.15.-q; 04.20.-q; 03.30.+p}
\vspace*{1.0cm}

\centerline{\large \bf Isotopic Grand Unification with the Inclusion of
Gravity}

\vspace*{0.5cm}
\centerline{R.M. Santilli}
\centerline{Institute for Basic Research}
\centerline{Palm Harbor, FL 34682-1577}
\centerline{ibr@gte.net; http://home1.gte.net/ibr/}

\vspace*{0.5cm}
\begin{abstract}
\noindent{\bf Abstract\/}: We introduce a dual lifting of unified gauge
theories, the first characterized by the {\it isotopies}, which are
axiom-
preserving maps into broader structures with positive-definite
generalized
units used for the representation of matter under the isotopies of the
Poincare' symmetry,
and the second characterized by the {\it isodualities},
which are anti-isomorphic maps with negative-definite generalized units
used for the representation of antimatter under the isodualities of the
Poincare' symmetry. We then submit, apparently for the
first time, a novel grand unification with the inclusion of gravity
for matter embedded in the generalized positive-definite units of
unified gauge theories
while gravity for antimatter is embedded in the isodual isounit.
We then show that the proposed grand
unification provides realistic possibilities for a resolution of the
axiomatic incompatibilities between gravitation and electroweak
interactions
due to curvature, antimatter and the fundamental space-time symmetries.
\end{abstract}
\vspace*{1.0cm}
\noindent{\large \bf 1. Introduction}
\smallskip

\par
\noindent In this note we study the structural incompatibilities for an
axiomatically consistent inclusion of gravitation $^{(1)}$ in the
unified gauge theories of electroweak interactions$^{(2)}$ due to:

{\bf (1) Curvature.} In fact, electroweak theories are essentially
structured on {\it Minkowskian} axioms, while gravitational theories are
formulated via {\it Riemannian} axioms, a disparity which is magnified
at
the operator level because of known technical difficulties of
quantum gravity$^{(3)}$, e.g., to provide a PCT theorem comparable
to that of electroweak interactions.

{\bf (2) Antimatter.} In fact, electroweak theories are {\it bona fide}
relativistic theories, thus characterizing antimatter via {\it
negative-energy}
solutions, while gravitation characterizes antimatter via
{\it positive-definite} energy-momentum tensors.

{\bf (3) Fundamental space-time symmetries.} In fact, the electroweak
interactions are based on the axioms of the special relativity,
thus verifying the fundamental {\it Poincare' symmetry} $P(3.1)$, while
such a basic symmetry is absent in contemporary gravitation.

Without any claim of uniqueness (see, e.g., the recent studies
on unified theories of monograph$^{(2m)}$ and references quoted
therein),
we present apparently for the first time a conceivable
resolution of the above  structural incompatibilities via the use of
the following new methods:

{\bf (A) Isotopies.} A baffling aspect in the inclusion
of gravity in unified gauge theories is their apparent geometric
incompatibility despite their individual beauty and experimental
verifications.

The view we would like to convey is that the above structural
incompatibility {\it is not} necessarily due
to insufficiencies in Einstein's field
equations, but rather to {\it insufficiencies in their mathematical
treatment}.
Stated in plain language, we believe that the achievement of axiomatic
compatibility between gravitation and electroweak interactions requires
a basically new mathematics, that is, basically new numbers, new
spaces, etc.

In the hope of resolving in due time this first structural
incompatibility,
Santilli$^{(4a)}$ proposed back in 1978 when at the Department of
Mathematics of
Harvard University under DOE support, a new mathematics based on the
so-called {\it isotopies}, and today known as {\it isomathematics},
which
was then studied in numerous works$^{(4-11)}$ (see in particular
the latest memoir$^{(4e)}$).

The isotopies are nowadays referred to {\it maps (also called liftings)
of any
given linear, local and canonical or unitary theory into its most
general
possible nonlinear, nonlocal and noncanonical or nonunitary extensions,
which are nevertheless capable of reconstructing linearity, locality and
canonicity or unitarity on certain generalized spaces and fields, called
isospaces and isofields.} From their Greek meaning, isotopies are
therefore
"axiom-preserving".

The fundamental isotopy of this note is that of the 4-dimensional unit
I = Diag. (1, 1, 1, 1) of
the Minkowskian and Riemannian space-times into a 4x4-dimensional,
everywhere invertible, Hermitean and positive-definite matrix ${\hat I}$
whose
elements have an arbitrary functional dependence on the local space-time
coordinates x, as well as any other needed variable,

$$ I = Diag. (1, 1, 1,1) \rightarrow {\hat I}(x, ...) = ({\hat
I}^{\mu}_{\nu}(x, ...)) =
{\hat I}^\dagger  =  [{\hat T}(x, ...)]^{-1} > 0,
\eqno(1)$$

\noindent with corresponding lifting of the conventional associative
product

$$ A\times B \rightarrow A{\hat{\times}}B = A\times {\hat T}\times B,
\eqno(2)$$

\noindent under which ${\hat I}(x, ...) = {[{\hat T}(x, ...)]}^{-1}$
is the correct left and right unit of the
new theory called {\it isounit}, in which case ${\hat T}(x, ...)$ is
called
the {\it isotopic element}.

When applicable, liftings (1) and (2) require for consistency the
reconstruction of {\it all} mathematical methods of contemporary
physics,
with no exception known to this author. In fact, they require new
numbers
and fields called {\it isonumbers} and {\it isofields} with an arbitrary
(positive-definite) unit, new spaces over isofields called {\it
isospaces},
the new {\it isodifferential calculus}, the new {\it isoeuclidean,
isominkowskian and isosymplectic geometries}, etc. which we cannot
possibly review in this note and must assume as known (see their latest
formulation in memoir$^{(4e)}$).

In a communication at the {\it VII Marcel Grossmann Meeting on General
Relativity} (mg7) held in 1994 at Stanford University, Santilli$^{(5a)}$
showed that
the isomathematics permits a novel classical and operator treatment of
gravitation which, on one side, preserves {\it Riemannian} metrics,
Einstein's
field equations and related experimental verifications while, on the
other
side, verifies the abstract {\it Minkowskian} axioms.

The above reformulation is evidently fundamental for this note, inasmuch
as
it offers realistic possibilities of resolving the structural
incompatibility
between electroweak and gravitational interactions due to curvature, by
reducing the latter to the axiomatic structure of the former {\it for
the case
of matter only} (see below for antimatter).

The main mechanism is that first presented at mg7 (Ref.$^{(5a)}$, p.
501) which
is based on the factorization of any given Riemannian metric (e.g.,
Schwarzschild metric$^{(1c)}$) g(x) into the Minkowski metric
$\eta(+1, +1, +1, -1)$

$$ g(x) = T(x)\times \eta,
\eqno(3)$$

\noindent where the {\it gravitational isotopic element} $T(x)$ is
evidently a
$4\time4$-dimensional matrix which is always
positive-definite from the locally Minkowskian character of Riemann.
The entire theory must then be reconstructed with respect to the
{\it gravitational isounit}

$$ {\hat I} = [{\hat T}(x)]^{-1} = \eta\times [g(x)]^{-1} > 0,
\eqno(4)$$

Note that the component truly representing curvature in the Riemannian
geometry {\it is not} the Riemannian metric g(x) but rather its isotopic

omponent T(x), trivially, because the remaining component $\eta$ is
flat.
It is then easy to see that {\it the isotopic treatment of gravity
formally eliminates
curvature, thus rendering gravitation axiomatically compatible with the
electroweak interactions}. In fact, curvature exists when the
gravitational
isotopic element T(x) is referred to the conventional space-time unit I,
while
curvature formally disappears when $T(x)$ is referred to a generalized
unit
which is its {\it inverse} $[T(x)]^{-1}$.

Reformulations (3) and (4) also imply the birth of a novel geometry, the
{\it isominkowskian geometry}, first submitted by Santilli$^{(6a)}$ in
1983
which, in more recent studies$^{(5e)}$ has resulted to in a symbiotic
unification of
the Minkowskian and Riemannian geometries, because it verifies all the
abstract axioms of the former, while preserving the machinery of the
latter (covariant derivatives, connections, etc.). The formulation of
gravity based on Eq.s (3) and (4) is then called {\it isominkowskian
gravity}.

Allow us to stress for clarity that we are here referring to a mere
{\it mathematical} reformulation of Einstein's historical field
equations
on the {\it mathematical} isominkowskian spaces (i.e., refer them to a
new unit ${\hat I}$) because the projection of the treatment
into the conventional space-tine
(i.e., when referred to the conventional space-time unit I)  recovers
the
said
historical equations in their totality. The above occurrences therefore
offer realistic hopes of resolving the baffling occurrence indicated
earlier,
i.e., the apparent incompatibility between Einstein's majestic
conception of
gravitation with the geometric structure of electroweak interactions.

The reader should be aware that the proposed resolution works
best where it is needed most, at the {\it operator} level.
In fact, Santilli$^{(5a)}$
showed at mg7 that {\it the operator formulation of the isominkowskian
representation of gravity verifies all abstract axioms and physical laws
of conventional "relativistic" quantum mechanics} (RQM). The emerging
new
theory is called {\it operator isogravity} (OIG) and merely consists
in embedding gravity in the {\it unit} of RQM.

The reader should be aware that the above classical and operator
isotopies
are supported by two, hitherto unknown symmetries, first presented in
memoir$^{(4f)}$ under the tentative name of
{\it isoselfscalar symmetries}, which are characterized by the
transforms

$$ \eta \rightarrow {\hat\eta} = n^{-2}\times \eta,
 I \rightarrow {\hat I} = n^2\times I,
\eqno(5)$$

\noindent where n is a parameter, and yield the symmetry
of the conventional Minkowskian interval

$$x^2 = (x^{\mu}\times\eta_{\mu\nu}\times x^{\nu})\times I =
 (x^{\mu}\times{\hat \eta}_{\mu\nu}\times x^{\nu})\times {\hat I} =
x^{\hat 2},
\eqno(6)$$

\noindent with a corresponding invariance for the Hilbert space

$$ < \phi | \times | \psi>\times I  =
< \phi | \times n^{-2}\times | \psi>\times (n^2\times I) =
< \phi | {\hat {\times}} | \psi>\times {\hat I}.
\eqno(7)$$

The isominkowskian representation of gravity then emerges from the above
classical and quantum symmetries via the axiom-preserving addition of an
x-dependence in the n-parameter, much along the transition from Abelian
to non-Abelian gauge theories.

{\bf (B) Isodualities.} Structural incompatibility (2) is only the
symptom of deeper problems in the contemporary treatment of antimatter.
To begin, matter is treated nowadays at {\it all} levels, from Newtonian
to electroweak interactions, while antimatter is treated only at the
level
of {\it second quantization.} Since there are serious indications that
half of the
universe could well be made up of antimatter, it is evident that a more
effective theory of antimatter must also apply at {\it all} levels.

At any rate, recall that charge conjugation in quantum mechanics is an
{\it anti-automorphic map}. As a result, no classical theory of
antimatter
can be axiomatically consistent via the mere change of the sign of the
charge, because it must be an anti-automorphic (or, more generally,
anti-isomorphic) image of that of matter (the alternative
classical formulation of antimatter of Ref. [12c] has been recently
brought
to the author's attention).

The current dramatic disparity in the treatment of matter and antimatter
also has its predictable problematic aspects.
Since we currently use only one type of quantization (whether naive of
symplectic), it is easy to see that {\it the operator image of the
contemporary
treatment of antimatter is not the correct charge conjugate state, but
merely a conventional state of particles with a reversed sign of the
charge}.

 The view here submitted is that, as it is the case for curvature,
the resolution of the above general shortcomings, including the
achievement of
compatibility in the treatment of antimatter between electroweak and
gravitational interactions, requires a basically novel mathematics.

Santilli therefore entered into a second search for another
novel mathematics under the uncompromisable condition of being an
anti-isomorphic image of the preceding isomathematics. After inspecting
a
number of alternatives, this author$^{(6c)}$ submitted in 1985
the following map of an arbitrary quantity Q (i.e., a number, or
a vector field or an operator) under the tentative name of {\it
isoduality}

$$Q \rightarrow Q^d = - Q^\dagger.
\eqno(8)$$

When applied to the {\it totality} of quantities and their
operations of a given theory of matter,
map (8) yields an anti-isomorphic image, as
axiomatically
needed for antimatter. Moreover, while charge conjugation is solely
applicable within operator settings, isoduality (8) is applicable at
{\it all} levels of study, beginning at the {\it Newtonian} level.

It is evident that map (8) implies
a new mathematics, that with negative units
called {\it isodual mathematics}$^{(7)}$, which includes
new numbers, new spaces, new calculus, etc. In reality we have two
different
isodual mathematics, the first is the anti-isomorphic image of the
conventional mathematics used for matter, and the second is the
anti-isomorphic
image of the preceding isomathematics.

The above characteristics have permitted the construction of the
novel {\it isodual theory of antimatter}$^{(7)}$
 which is equivalent,
although anti-isomorphic, to that of matter, and which therefore begins
at the primitive {\it Newtonian} level and then continues at the
analytic
and quantum levels, in which case it results in equivalence to charge
conjugation for massive particles (see later on for photons).

Most importantly, the isodual theory of antimatter has resulted in
agreement with all available  classical and quantum experimental data
on antimatter.

It is evident that isodualities offer a realistic possibility of
resolving
the second structural problem between electroweak and gravitational
interactions because antimatter can be treated in both cases with
{\it negative-energy}.
This is due to the fact that isodualities imply the transition from
the conventional space-time units of matter $I=Diag. (1, 1, 1, 1) > 0$
to their
negative image $I^d = - I < 0$. As a result,
{\it all characteristics of matter change sign in the
transition to antimatter under isoduality}, thus yielding
the correct conjugation of charge, as well as negative energy,
negative energy-momentum tensor, and, inevitably, negative time.
The historical
objections against these negative values are inapplicable,
because they are tacitly
referred to the conventional positive units. In fact, negative energy
and time
referred to negative units are fully equivalent, although
antiautomorphic,
to the conventional positive energy and time referred to positive units.

The reader should also be aware that the isodual theory of antimatter
was
born from properties of the conventional Dirac equation

$$[\gamma^\mu\times (p_\mu-e\times A_\mu/c)+i\times
m]\times\Psi(x)=0,\eqno(9a)$$

$$\gamma^k=\left( \begin{array}{cc}
0 & \sigma^k \\
-\sigma^k & 0
\end{array} \right),\eqno(9b) $$ $$ \gamma^4=i\times\left(
\begin{array}{cc}
I_s & 0. \\
0 & -I_s \end{array} \right)
\eqno(9c)$$

In fact, as one can see, {\it the negative
unit} $I^d_s = Diag. (-1, -1)$
{\it appears in the very structure of} $\gamma_4$.
The isodual theory was then constructed precisely around Dirac's unit
$I^d_s$

In essence, Dirac assumed that the negative-energy solutions of his
historical equation behaved in an unphysical way because tacitly
referred to
the conventional mathematics of his time, that with {\it positive units}
$I_s > 0$.
Santilli$^{(7)}$ showed that, when the same negative-energy
solutions are referred to
the {\it negative units} $I^d_s < 0$, they behaved in a fully physical
way. This
eliminates the need of second quantization for the treatment of
antiparticles
(as expected in a theory of antimatter beginning at the Newtonian
level),
and permits the reformulation of the equation in the form

$$[\tilde{\gamma}^\mu\times(p_\mu-e\times A/c)+i\times
m]\times\tilde{\Psi}(x) = 0,
\eqno(10a)$$

$$\tilde{\gamma}_k=\left( \begin{array}{cc}
0 & \sigma^d_k \\
\sigma_k & 0 \end{array} \right), \; \;
\tilde{\gamma}^4=i\left( \begin{array}{cc}
I_s & 0, \\
0 & I_s^d \end{array} \right),
\eqno(10b)$$

$$\{\tilde{\gamma}_\mu,\tilde{\gamma}_\nu\}=2\eta_{\mu\nu}, \; \;
\tilde{\Psi}=-\tilde{\gamma}_4\times\Psi=i\times
\left( \begin{array}{c}
\Phi \\ \Phi^d \end{array} \right),
\eqno(10c)$$

\noindent where $\Phi(x)$ is now two-dimensional, which
is fully symmtrized between particles and antiparticles.

As was the case for the preceding isotopies,
the isodual theory of antimatter also sees its solid roots in
two additional novel symmetries, also unknown until recently, and first
presented
in memoir$^{(4f)}$, the first holding for the conventional Minkowski
interval

$$x^2 = (x^{\mu}\times\eta_{\mu\nu}\times x^{\nu})\times I =
[x^{\mu}\times (-n^{-2}\times \eta_{\mu\nu})\times x^{\nu}]\times
(-n^2\times I) =$$

$$= (x^{\mu}\times {\hat \eta}^d_{\mu\nu}\times x^{\nu})\times {\hat
I}^d =
x^{d2d},
\eqno(11)$$

\noindent and the second holding for the Hilbert space

$$ < \phi | \times | \psi>\times I  =
< \phi | \times (-n^{-2})\times | \psi>\times (-n^2\times I) =
< \phi | \times {\hat T}^d\times | \psi>\times {\hat I}^d,
\eqno(12)$$

\noindent which ensure that all physical laws for matter
also hold for antiparticles under our isodual representation, with
corresponding symmetries for the isodual expressions.

The axiom-preserving lifting of the
n-parameter to an explicit x-dependence then yields the
 {\it isodual isominkowskian treatment of gravity for antimatter}
with basic structures,

$$ g(x) = {\hat T}(x)\times \eta \rightarrow g^d(x) = -g(x) =
{\hat T}^d(x) \times^d \eta^d, \eta \rightarrow \eta^d = - \eta,
\eqno(13a) $$

$${\hat I}(x) = [{\hat T}(x)]^{-1} \rightarrow {\hat I^d(x)} =
[{\hat T}^d(x)]^{-1}.
\eqno(13b)$$

        As we shall see in the next section rules (3)-(4) and (13) can
indeed
be
implemented within unified gauge theories

{\bf (C) Isotopies of the Poincare' symmetry and their isoduals.}
Judging from
the studies herein reported, the most severe problems
of compatibility between gravitation and electroweak interactions for
both matter and antimatter appeared precisely were expected, in the {\it
fundamental space-time symmetries}, because of the disparity indicated
earlier
of the validity of the Poincare' symmetry for electroweak interactions
and
its absence for gravitation.

The latter problems called for a third series of studies presented in
Ref.s$^{(6)}$ on the {\it isotopies and isodualities
of the Poincare' symmetry} ${\hat P}(3.1)$, today called the
{\it Poincare'-Santilli isosymmetry and its isodual}$^{(8b-8e)}$,
which include the isotopies and isodualities of:
the rotational symmetry$^{(6c)}$; the Lorentz symmetry in
classical$^{(6a)}$ and
operator version$^{(6b)}$; the SU(2)-spin symmetry$^{(6d)}$; the
Poincare' symmetry$^{(6e)}$;, and the spinorial covering of the
Poincare' symmetry$^{(6f)}$ (see monographs$^{(6g)}$ for comprehensive
studies).

We are here referring to the reconstruction of the conventional
symmetries with
respect to an arbitrary positive-definite unit (1), for the isotopies,
and with respect to
an arbitrary negative-definite unit, for the isodualities. This
reconstruction
yields the most general known nonlinear, nonlocal and noncanonical
liftings
of conventional symmetries, while being locally isomorphic (for
isotopies) or
anti-isomorphic (for isodualities) to the original symmetries.

One should be aware that the above structures required the prior
step-by-step isotopies and isodualities of Lie's theory (enveloping
associative algebras, Lie algebras, Lie groups, transformation and
representation theories, etc.), originally proposed
by Santilli$^{(4a)}$ in 1978, studied in numerous subsequent works (see
monographs$^{(4c,6g)}$) and today called the {\it Lie-Santilli isotheory
and its isodual}$^{(8-10)}$.

It is evident that the isopoincare' symmetry and its isodual have
fundamental
character for this note. In fact, one of their primary applications has
been the achievement of the universal {\it symmetry} (rather than
covariance) of all possible
Riemannian line elements in their isominkowskian representation$^{(6)}$.
Once the unit of gauge theories is lifted to represent gravitation,
electroweak interactions will also obey the isopoincare' symmetry for
matter and its isodual for antimatter, thus offering realistic hopes for
the resolution of the most difficult problem of compatibility, that
for space-time symmetries.

Perhaps unexpectedly, the fundamental space-time symmetry of the
grand unified theory inclusive of gravitation submitted in this note
is the total symmetry of the {\it conventional} Dirac equation,
here written with their underlying spaces and units

$$S_{Tot}  = \{SL(2.C)\times T(3.1)\}\times
\{SL^d(2.C^d)\times^dT^d(3.1)\},\eqno(14a)$$

$$M_{Tot}  =  \{M(x,\eta,R)\times
S_{spin}\}\times\{M^d(x^d,\eta^d,R^d)\times^dS^d_{spin}\}
\nonumber,\eqno(14b)$$

$$I_{Tot}  = \{I_{orb}\times I_{spin}\}\times
\{I^d_{orb}\times^dI^d_{spin}\}, \nonumber
\eqno(14c)$$

\noindent which has recently emerged as being {\it twenty-two
dimensional}.

To see the above occurrence, the reader should be aware that
isodualities
imply yet another new symmetry called {\it isoselfduality}$^{7)}$, which
is
the invariance under the isodual map (8). Dirac's gamma matrices
verify indeed this new symmetry (from which the symmetry itself was
derived
in the first place), i.e.,
$\gamma_{\mu}\rightarrow {\gamma}^d_{\mu} =
 -{\gamma}^{\dagger}_{\mu} = {\gamma}_{\mu}$.
As a result, contrary to a popular belief throughout this century, the
Poincare' symmetry {\it cannot} be the total symmetry of Dirac's
equations,
evidently because it is not isoselfdual. For evident reasons of
consistency, the total symmetry of Eq.s (9) must
also be isoselfdual as the gamma-matrices are. This resulted in the
identification of the total symmetry (14a) which is indeed
isoselfdual.

To understand the dimensionality of symmetry one must first
recall that isodual spaces are independent from conventional spaces. The
doubling of the conventional dimensionality then  yields {\it twenty}
dimensions. The additional two dimensions are given by the novel
isoselfscalarity, i.e., invariance (6)-(7) and their isoduals (11)-12).

In short, the grand unification proposed in this note is based
on the axiomatic structure of the conventional Dirac's equations,
as emerged from the novel insights of memoir$^{(4f)}$,
and merely subjected to axiom-preserving liftings, in which
the inclusion of gravitation for matter
is permitted by the novel isoselfscalar symmetries
(6)-(7), and that for antimatter by the anti-isomorphic images
(11)-(12).

The reader should not be surprised that the four new invariances
(6)-(7) and (11)-(12)
remained undetected throughout this century. In fact, their
identification
required the prior discovery of {\it new numbers}, first the numbers
with arbitrary positive units for invariances (6)-(7), and
then the additional new  numbers with
arbitrary negative units for invariances (11)-(12).

\vspace*{0.5cm}
\noindent {\Large \bf 2. Isotopic Gauge Theory}
\vspace*{0.5cm}

The isotopies of gauge theories were first studied in
1980's by Gasperini $^{(11a)}$,
followed by
Nishioka$^{(10b)}$, Karajannis and Jannussis$^{(11c)}$ and others,
and ignored thereafter.
These studies were defined on conventional spaces over
conventional fields and via the conventional differential calculus. As
such,
they are not invariant, as we learned only recently in memoirs$^{(4f)}$.

In this section we shall introduce, apparently for the first time, the
{\it isotopies of gauge theories}, or {\it isogauge theories} for short,
formulated in an invariant way, that is, on
isospaces over isofields and characterized by the isodifferential
calculus
of memoir $^{(4e)}$. The {\it isodual isogauge theories} are apparently
introduced
in this note for the first time.

The essential mathematical methods needed for an axiomatically
consistent and invariant formulation of the isogauge theories are the
following:

{\bf (1) Isofields}$^{(4d)}$ of isoreal numbers
${\hat R}({\hat n},{\hat +},{\hat {\times}})$ and  isocomplex numbers
${\hat C}({\hat c},{\hat +},{\hat {\times}})$
with: additive isounit ${\hat 0} = 0$;
generalized multiplicative isounit ${\hat I}$ given by Eq. (1);
elements, isosum, isoproduct and related
generalized operations,
$$ {\hat a} = a\times {\hat I}, a = n, c,
{\hat a}{\hat +}{\hat b} = (a + b)\times {\hat I},
{\hat a}{\hat {\times}}{\hat b} = {\hat a}\times {\hat T}\times {\hat b}
= (a\times b)\times {\hat I},\eqno(15a)$$
$${\hat a}^{\hat n} = {\hat a}{\hat {\times}}{\hat a}{\hat {\times}} . .
. {\hat {\times}}{\hat a},
{\hat  a}^{\hat {1/2}} = a^{1/2}\times {\hat I}^{1/2},
{\hat a}{\hat /}{\hat b} = ({\hat a}/{\hat b})\times {\hat I}, etc.
\eqno(15b)$$

{\bf (2) Isominkowski spaces}$^{(6a)}$
${\hat M} = {\hat M}({\hat x},{\hat \eta},{\hat R})$
with isocoordinates ${\hat x} = x \times {\hat I} = \{x^\mu \} \times
{\hat I}$, isometric ${\hat N} ={\hat \eta}\times {\hat I} =
[{\hat T}(x, \dots) \times\eta]\times {\hat I}, $
and  {\it isointerval} over the isoreals ${\hat R}$

$$({\hat x} - {\hat y})^{\hat 2} = [({\hat x} - {\hat y}^{\mu}
{\hat {\times}}{\hat N}_{\mu\nu}{\hat{\times}}({\hat x} - {\hat
y})^{\nu} =
[(x - y)^{\mu}\times {\hat {\eta}}_{\mu\nu}\times (x - y)^{\nu}]\times
{\hat
I},
\eqno(16)$$

\noindent equipped with Kadeisvili isocontinuity$^{(10a)}$ and
Tsagas-Sourlas isotopology$^{(10b)}$ (see also Aslander and
Keles$^{(10d)}$).
A more technical formulation of the isogauge theory can be done via the
isobundle formalism on isogeometries recently reached by
Vacaru$^{(10e)}$,
which will be studied in a future work.

{\bf (3) Isodifferential calculus}$^{(4e)}$ characterized by the
following isodifferentials and isoderivatives

$${\hat d}{\hat x}^{\mu} = {\hat I}^{\mu}_{\nu}\times d{\hat x}^{\nu},
{\hat d}{\hat x}_{\mu} = {\hat T}_{\mu}^{\nu}\times {\hat x}_{\nu},
{\hat {\partial}}_{\mu}{\hat f} = {\hat {\partial}}{\hat f}{\hat /}{\hat
{\partial}}{\hat x}^{\mu} =
({\hat T}_{\mu}^{\nu}\times {\partial}_{\nu}f)\times {\hat I},
\eqno(17a)$$
$${\hat {\partial}}^{\mu}{\hat f} =
({\hat I}^{\mu}_{\nu}\times \partial_{\nu}f)\times {\hat I},
{\hat {\partial}}{\hat x}^{\mu}{\hat /}{\hat {\partial}}{\hat x}^{\nu} =
{\hat {\delta}}^{\mu}_{\nu} = {\delta}^{\mu}_{\nu}\times {\hat I}, etc.
\eqno(17b)$$

{\bf (4) Isofunctional isoanalysis}$^{(6g)}$, including the
reconstruction of
all conventional and special functions and transforms into a form
admitting
of ${\hat I}$ as the left and right unit. Since the isominkowskian
geometry
preserves the Minkowskian axioms, it allows the preservation of the
notions of
straight and intersecting lines, thus permitting the reconstruction of
trigonometric and hyperbolic functions under the Riemannian metric
$g(x) = {\hat {\eta}}^{(6g)}$.

{\bf (5) Isominkowskian geometry}$^{(5e)}$, i.e., the geometry of
isomanifolds ${\hat M}$ over the isoreals ${\hat R}$, which satisfies
all
abstract Minkowskian axioms because of the joint liftings
${\eta} \rightarrow {\hat{\eta}} = T(x, ... )\times {\eta}$
and $I\rightarrow {\hat I} = T^{-1}$, while preserving
the machinery of Riemannian spaces
(covariant derivatives, connections, etc.), although
expressed in terms of the isodifferential calculus for consistency. In
this new geometry {\it Riemannian} line elements are turned into
identical {\it Minkowskian} forms via the embedding of gravity in the
deferentials, e.g., for the Schwarzschild exterior metric we have
the isominkowskian reformulation
(Ref.[5e], Eq.s (2.57), where the space-time coordinates
are assumed to be covariant,

$${\hat d}{\hat s} = {\hat d}{\hat r}^{\hat 2} {\hat +}
{\hat r}^{\hat 2}{\hat {\times}}
({\hat d}{\hat {\theta}}^{\hat 2} {\hat +} isosin^{\hat 2}{\hat
{\theta}})
{\hat -}{\hat d}{\hat t}^{\hat 2},
\eqno(18a)$$

$${\hat d}{\hat r} = {\hat T}_r\times d{\hat r},
{\hat d}{\hat t} = {\hat T}_t\times d{\hat t},
{\hat T}_r = (1 - 2\times M/r)^{-1}, {\hat T}_t = 1 - 2\times M/r.
 \eqno(18b)$$

{\bf (6) Relativistic hadronic mechanics}$^{(4f)}$ characterized by
the {\it isohilbert space} ${\hat {\cal H}}$ first introduced by Myung
and
Santilli$^{(9e)}$ in 1982 with {\it isoinner product and
isonormalization}
over ${\hat C}$

$$<{\hat {\phi}} {\hat |} {\hat {\psi}}> =
<{\hat {\phi}}| {\hat {\times}}|{\hat {\psi}}>\times {\hat I},
<{\hat {\psi}} {\hat |} {\hat {\psi}}> = {\hat I}.
\eqno(19)$$

Among the various properties we
recall that: the {\it isohermiticity} on ${\hat {\cal H}}$
coincides with the conventional Hermiticity (thus, all conventional
observables
remain observables under isotopies); the isoeigenvalues of isohermitean
operators are real and conventional (because of the identities
${\hat H}{\hat {\times}}|{\hat {\psi}}> = {\hat E}{\hat {\times}}|{\hat
{\psi}}> = E\times|{\hat {\psi}}>)$; the condition of
{\it isounitarity} on ${\hat {\cal H}}$
over ${\hat C}$ is given by
${\hat U}{\hat {\times}}{\hat U}^{\dagger} =
{\hat U}^{\dagger}{\hat {\times}}{\hat U} = {\hat I}$
(see memoir$^{(4f)}$ for details).

{\bf (7) The Lie-Santilli isotheory}$^{(4,6,8d,10c)}$ with:
conventional (ordered) basis of generators $X = (X_k),$ and
parameters $w = (w_k)$  k = 1, 2, ..., n,
only formulated in isospaces over isofields with a
common isounit;
universal enveloping isoassociative algebras ${\hat {\xi}}$ with
infinite-dimensional basis characterized by the
isotopic Poincare'-Birkhoff-Witt theorem$^{(4a,4c,6g)}$

$$ {\hat I}, {\hat X}_i{\hat {\times}}{\hat X}_j, (i \leq j),
{\hat X}_i{\hat {\times}}{\hat X}_j\times{\hat X}_k,
(i\leq j \leq k;, . . .
\eqno(20)$$

\noindent Lie-Santilli isoalgebras {\it (loc. cit)}

$$[{\hat X}_i{\hat ,}{\hat X}_j] = {\hat X}_i{\hat {\times}}{\hat X}_j
-
{\hat X}_j {\hat {\times}}{\hat X}_i  =
{\hat C}_{ij}^k(x, ...){\hat {\times}} {\hat X}_k;
 \eqno(21) $$

\noindent where the ${\hat C}$'s are the structure isofunctions; and
 isogroups characterized by isoexponentiation
on ${\hat {\xi}}$ with structure {\it (loc. cit.)}

$${\hat e}^{\hat X} = {\hat I} {\hat +} {\hat X}{\hat /}{\hat 1}{\hat
!}
{\hat +} {\hat X}{\hat {\times}}{\hat X}{\hat /}{\hat 2}{\hat !} {\hat
+} . . . = ( e^{\hat X\times{\hat T}})\times {\hat I} =
{\hat I}\times ({e^{{\hat T}\times {\hat X}}}),
\eqno(22)$$

Despite the isomorhism between isotopic and conventional structures,
the lifting of Lie's theory is
nontrivial because of the appearance of the matrix ${\hat T}$ with
nonlinear elements in the very {\it exponent}
of the group structure, Eq.s (22). To avoid misrepresentations, one
must therefore keep in mind that the isotopies of Lie's theory {\it were
not}
proposed  to build "new algebras" (an impossible task since all simple
Lie
algebras
are known from Cartan's classification), but to construct instead the
most general possible nonlinear, nonlocal and noncanonical or nonunitary
"realizations" of known Lie algebras.

Another important aspect the reader should keep in mind is that the
isotopies
are such to reconstruct linearity, locality and canonicity or unitarity
on
isospaces over isofields, called {\it isolinearity, isolocality and
isocanonicity or isounitarity}. As a result, the use of
the conventional {\it linear} transformations on M over R, $X' =
A(a)\times x$
violates {\it isolinearity} on ${\hat M}$ over ${\hat R}$. In general,
{\it any}
use of conventional mathematics for isotopic theories leads to a number
of inconsistencies which generally remain undetected by nonexperts in
the field.

We are now minimally equipped to introduce the desired isogauge theory
which can be characterized by an
 $n$-dimensional connected and non-abelian isosymmetry ${\hat G}$
with: basic $n$-dimensional isounit (1); isohermitean operators
${\hat X}$ on an isohilbert space ${\hat {\cal H}}$ over the isofield
${\hat C}({\hat c},{\hat +},{\hat {\times}})$;
universal enveloping
associative algebra ${\hat {\xi}}$ with infinite isobasis (20);
isocommutation rules (21); isogroup structure

$$ {\hat U} = \, {\hat e}^
{- i\times X_k \times \theta(x)_k} =
 \, (e^{- i\times X_k \times {\hat T}\times \theta(x)_k})\times {\hat
I},
{\hat U}^\dagger {\hat {\times}}{\hat U} = {\hat I};
\eqno(23) $$

\noindent where one should note the appearance of the
gravitational isotopic elements in the exponent, and
the parameters $\theta(x)_k$ now depend on the
isominkowski space; isotransforms of the isostates on
${\hat {\cal H}}$

$$ {\hat {\psi}}' = {\hat U} {\hat {\times}}{\hat {\psi}} =
\bigl(e^{-i \times X_k \times {\hat T}(x, ...)
\times \theta(x)_k}\bigr)\times {\hat {\psi}};
\eqno(24)  $$

\noindent isocovariant derivatives$^{(5e)}$

$$ {\hat D}_\mu {\hat {\psi}} = ({\hat {\partial}}_\mu - i
{\hat {\times}}{\hat g} {\hat {\times}}{\hat A}({\hat x})_{\mu}^k
{\hat {\times}}{\hat  X}_k){\hat {\times}}{\hat {\psi}};
\eqno(25) $$

\noindent iso-Jacobi identity

$$ [{\hat D}_\alpha{\hat ,}[{\hat D}_\beta{\hat ,} {\hat D}_\gamma]]
{\hat +} [{\hat D}_\beta{\hat ,}[{\hat D}_\gamma{\hat ,} {\hat
D}_\alpha]]
{\hat +}[{\hat D}_\gamma{\hat ,}[{\hat D}_\alpha{\hat ,} {\hat
D}_\beta]] = 0;
\eqno(26)$$

\noindent where $g$ and ${\hat g} = g\times {\hat I}$ are the
conventional and isotopic coupling constants, $A(x)_\mu^k \times X_k$
and ${\hat A}({\hat x})_\mu^k {\hat {\times}}{\hat X}_k =
[A(x)_\mu^k \times X_k]\times {\hat I}$
are the gauge and isogauge potentials;
isocovariance

$$({\hat D}_\mu{\hat {\psi}})' =
({\hat {\partial}}_{\mu} {\hat U}) {\hat {\times}} {\hat {\psi}}
{\hat +}{\hat U}{\hat {\times}}
({\hat {\partial}}_{\mu} {\hat {\psi}})
{\hat -}{\hat i}{\hat {\times}}{\hat g}{\hat {\times}}{\hat A}'
({\hat x})_{\mu} {\hat {\times}}{\hat  {\psi}} =
{\hat U}{\hat {\times}}{\hat D}_{\mu} {\hat {\psi}},
 \eqno(27a) $$
$${\hat A}({\hat x})'_{\mu} = -{\hat g}^{-{\hat 1}} {\hat {\times}}
 [{\hat {\partial}}_{\mu} {\hat U}({\hat x})]{\hat {\times}}
{\hat U}({\hat x})^{-{\hat 1}},
 \eqno(27b) $$
$${\hat {\delta}}{\hat A}({\hat x})_{\mu}^k = -{\hat g}^{-{\hat 1}}
{\hat {\times}}{\hat {\partial}}_{\mu} {\hat {\theta}}({\hat x})^k
{\hat +}{\hat  C}_{ij}^k {\hat {\times}} {\hat {\theta}}({\hat x})^i
{\hat {\times}}{\hat A}({\hat x})_{\mu}^j,
 \eqno(27c) $$
$${\hat {\delta}}{\hat {\psi}} = -{\hat i}{\hat {\times}}{\hat g}
{\hat {\times}}{\hat {\theta}}({\hat x})^k {\hat {\times}}{\hat X}_k
{\hat {\times}}{\hat {\psi}};
 \eqno(27d) $$

\noindent non-abelian iso-Yang-Mills fields

$${\hat F}_{\mu\nu} = {\hat i}{\hat {\times}}{\hat g}^{-{\hat 1}}
{\hat {\times}} [{\hat D}_{\hat {\mu}}, {\hat D}_{\nu}] {\hat {\psi}},
\eqno(28a)$$
$${\hat F}_{\mu\nu}^k = {\hat {\partial}}_{\mu} {\hat A}_{\nu}^k
{\hat -}{\hat {\partial}}_{\nu} {\hat A}_{\mu}^k
{\hat +}{\hat g}{\hat {\times}}{\hat C}_{ij}^k
{\hat {\times}}{\hat A}_{\mu}^i {\hat {\times}}{\hat  A}_{\nu}^j ;
\eqno(28b)
$$

\noindent related isocovariance properties

$${\hat F}_{\mu\nu} \rightarrow{\hat  F}_{\mu\nu}' =
{\hat U}{\hat {\times}}{\hat F}_{\mu\nu} {\hat {\times}}{\hat U}^{-1},
\eqno(29a) $$
$$Isotr ({\hat F}_{{\mu\nu}'} {\hat {\times}}{\hat F}^{{\mu\nu}'}) =
Isotr  ({\hat F}_{\mu\nu} {\hat {\times}}{\hat F}^{\mu\nu}),
\eqno(29b) $$
$$[{\hat D}_\alpha{\hat ,} {\hat F}_{\beta\gamma}]
{\hat +} [{\hat D}_\beta{\hat ,} {\hat F}_{\gamma\alpha}]
{\hat +} [{\hat D}_\gamma{\hat ,} {\hat F}_{\alpha `\beta}] \equiv 0;
\eqno(29c) $$

\noindent derivability from the isoaction

$${\hat S}= {\hat {\int}}{\hat d}^{\hat 4}{\hat x}
(-{\hat F}_{\mu\nu} {\hat {\times}}{\hat F}^{\mu\nu}{\hat /}{\hat 4}) =
{\hat {\int}}{\hat d}^{\hat 4}{\hat x}
(-{\hat F}_{\mu\nu}^k {\hat {\times}}{\hat F}^{\mu\nu}_k {\hat /}{\hat
4});
  \eqno(30) $$

\noindent where ${\hat {\int}} = {\int {\times {\hat I}}}$,
plus all other familiar properties in isotopic formulation.

The {\it isodual isogauge theory} is the image of the preceding theory
following the application of the isodual map (8) to the totality of
quantities and their operations.
The latter theory is characterized by the isodual isogroup ${\hat G}^d$
with isodual isounit ${\hat I}^d = -{\hat I}^{\dagger} =  - {\hat I}$.
The base fields are the field
${\hat R}^d({\hat n}^d,{\hat +}^d,{\hat {\times}}^d)$ of isodual isoreal
numbers ${\hat n}^d = -{\hat n} = -n\times {\hat I}$ and the field
${\hat C}^d({\hat c}^d,{\hat +}^d,{\hat {\times}}^d)$ of
isodual isocomplex numbers ${\hat c}^d =
-(c\times {\hat I})^{\dagger} =
(n_1 - i\times n_2)\times {\hat I}^d = (-n_1 + i\times n_2)\times {\hat
I}$.

The
carrier spaces are the isodual isominkowski space
${\hat M}^d({\hat x}^d,{\hat {\eta}}^d,{\hat R}^d)$ on ${\hat R}^d$ and
the isodual isohilbert space ${\ {\cal H}}^d$ on ${\hat C}^d$ with
isodual isostates $|{\hat {\psi}}>^d = -|{\hat {\psi}}>^{\dagger}$ and
isodual isoinner product $<{\hat {\phi}}|^d\times {\hat T}^d\times
|{\hat {\psi}}>^d\times {\hat I}^d$. It is instructive to verify that
all eigenvalues of isodual isohermitean operators are ${\it
negative-definite}$
(when projected in our space-time),
${\hat H}^d\hat {\times}^d|{\hat {\psi}}>^d =
(-E)\times |{\hat {\psi}}>$.

${\hat G}^d$ is characterized by
the isodual Lie-Santilli isotheory with
isodual generators ${\hat X}^d = -{\hat X}$, isodual isoassociative
product
${\hat A}^d\hat{\times}^d{\hat B}^d = {\hat A}^d\times {\hat T}^d\times
{\hat B}^d,
{\hat T}^d = - {\hat T}$ and related isodual isoenveloping and
Lie-Santilli
isoalgebra. The elements of ${\hat G}^d$ are the isodual isounitary
isooperators ${\hat U}^d({\hat {\theta}}^d({\hat x}^d)) =
- {\hat U}(-{\hat {\theta}}(-{\hat x}))$. In this way, the isodual
isogauge theory is seen to be an anti-isomorphic image of the
preceding theory, as desired.

It is an instructive exercise for the reader interested in learning
the new techniques to study first the isodualities of the {\it
conventional} gauge
theory (rather than of their isotopies), and show that they essentially
provide a mere reinterpretation of the usually discarded, advanced
solutions
as characterizing antiparticles. Therefore, in the isoselfdual theory
with total gauge symmetry ${\hat G}\times{\hat G}^d$, isotopic retarded
solutions are associated with particles and advanced isodual solutions
are
associated with antiparticles.

No numerical difference is expected in the above reformulation because
in conventional theories
particles and antiparticles are represented with retarded
solutions while advanced solutions are generally discarded. By
comparison,
in the isodual theory retarded solutions are solely used for particles
and
advanced solutions are solely used for antiparticles,
the two solutions being formulated in their respective different
spaces over different fields.

It is also recommendable for the interested reader to verify that the
isotopies
are indeed equivalent to charge conjugation for all massive particles,
with the exception of the photon$^{(7c)}$. In fact, isodual theories
predict
that the antihydrogen atom$^{(12b)}$ emits a new photon, tentatively
called by this author the {\it isodual photon}, which coincides
with the conventional photon for all possible interactions,
thus including electroweak interactions, {\it except
gravitation}$^{(7c)}$.
This indicates that the isodual map is inclusive of charge conjugation
for massive particles, but it is broader than the latter.

Isodual theories in general, thus including the proposed grand
unification,
predict that all {\it stable} isodual particles,
such as the isodual photon, the isodual electron (positron), the
isodual proton (antiproton) and their bound states
(such as the antihydrogen atom),
experience {\it antigravity}
in the field of the Earth (defined as the reversal of the sign of the
curvature
tensor). If confirmed, the prediction may offer the possibility in the
future
to ascertain whether far away galaxies and
quasars are made-up of matter or of antimatter

Known objections against antigravity are inapplicable because they
are tacitly referred to positive units and also because the isodual
theory
predicts that particle-antiparticle bound states such as the
positronium,
experience attraction in both fields of matter and antimatter$^{(7)}$.
The
latter predictions are currently under experimental study
by Mills$^{(11a)}$ and others$^{(11b)}$.

We also note that the isotopies leave unrestricted the functional
dependence of the isounit (1), provided that it is positive-definite. In
this note we use only the x-dependence to represent {\it exterior}
gravitational problems in vacuum. The isotheory also admits an arbitrary
nonlinearity in the velocities and other variables which is used for
the study of ${\it interior}$ gravitational problems. The
isotheory
naturally admits a dependence of the isounit on the {\it wavefunctions}
and their derivatives while preserving isolinearity in isospace (thus
preserving the superposition principle, as needed for a consistent
representation of composite systems). For these and
other aspects we refer the reader to memoir$^{(4f)}$.

We finally note that the isomathematics is a particular cases of the
broader {\it genomathematics}$^{(4a,4e,4f)}$, which occurs for
non-Hermitean generalized units and is used for an axiomatization
of irreversibility. In turn, the genomathematics is a particular case
of the {\it hypemathematics}$^{(4e,4f)}$, which occurs
when the generalized units are given by ordered {\it sets} of
non-Hermitean
quantities and is used for the representation of multivalued
complex systems (e.g.,
biological) in irreversible conditions. Evidently both the
genomathematics and hypermathematics admit an anti-isomorphic image
under isoduality (an outline of these novel mathematics can be found in
Page 18 of Web Site$^{(9y)}$).

In conclusion the methods outlined in this note permit the study
of {\it seven} liftings of conventional gauge theories;
(1) the {\it isodual gauge
theories} for the treatment of antimatter without gravitation in vacuum;
(2,3), the {\it isogauge theories and their isoduals}, for the inclusion
of
gravity for matter and antimatter in reversible conditions in vacuum
(exterior gravitational problem);
(4,5) the
{\it genogauge theories and their isoduals}, for the inclusion of
gravity
for matter and antimatter in irreversible interior conditions
(interior gravitational problems); and
(6,7) the {\it hypergauge theories and
their isoduals}, for multivalued and irreversible generalizations.
This note is restricted to theories (1, 2, 3).

\vspace*{0.5cm}
\noindent {\Large \bf 3. Iso-Grand-Unification}
\vspace*{0.5cm}

In this note we have submitted, apparently for the first time,
an Iso-Grand-Unification (IGU) with the inclusion of gravity
characterized by the total isoselfdual symmetry

$${\hat S}_{Tot} = ({\hat {\cal P}}(3.1){\hat {\times}}{\hat G})\times
({\hat {\cal P}}(3.1)^d {\hat {\times}}^d{\hat G}^d) = $$

$$= [{\hat {SL}}(2,{\hat C}){\hat {\times}}{\hat T}(3.1)]\times
[{\hat {SL}}^d(2,{\hat C}^d){\hat {\times}}^d{\hat T}^d(3.1)],
\eqno(31) $$

\noindent where ${\hat {\cal P}}$ is the
Poincare'-Santilli isosymmetry$^{(10c)}$ in its
isospinorial realization$^{(6f)}$, ${\hat G}$ is the isogauge symmetry
of the preceding section and the remaining structures are the
corresponding
isoduals.

Without any claim of a final solution, it appears that the proposed IGU
does indeed offer realistic possibilities of at least
resolving the axiomatic incompatibilities (1), (2) and (3) between
gravitational and electroweak interactions indicated in Sect. 1. In
fact,
IGU represents gravitation
in a form geometrically compatible with that of the
electroweak interactions, represents  antimatter at all
levels via negative-energy solutions,
and characterizes both gravitation as well as
electroweak interactions via the universal isopoincare' symmetry.

It should be indicated
that we are referring here to the ${\it axiomatic}$ consistency.
The ${\it physical}$
consistency is a separate problem which
cannot possibly be investigate in this introductory note and will be
investigated in future works. At this point we merely mention the
general
rule according to which isotopic liftings preserve not only the original
axioms, but also the original numerical values$^{(6g)}$ (as an
example, the image in isominkowskian space over the isoreals of the
light
cone, not only is a perfect cone, but a cone with the original
characteristic angle, thus preserving the speed of light in vacuum
as the maximal causal speed in isominkowskian space). This occurrence
provides
realistic hopes for the joint achievement of
axiomatic and physical consistency.

The reader should be aware that the methods of
the recent memoir$^{(4f)}$ permit a truly elementary,
explicit construction of the proposed IGU.
As well known, the transition from the Minkowskian
metric ${\eta}$ to Riemannian metrics $g(x)$ is a {\it noncanonical
transform}
at the classical level, and,
therefore, a {\it nonunitary
transform} at the operator level. The
method herein considered for turning a gauge theory into an IGU
consists in the following representation of the selected gravitational
model, e.g., Schwarzschild's model,

$$g(x) = T(x)\times {\eta}, T(x) = (U\times U^{\dagger})^{-1},
\eqno(32a) $$
$$ U\times U^{\dagger}) = Diag. [(1 - 2\times M/r)\times
Diag. (1, 1, 1), (1 - 2\times M/r)^{-1}),
\eqno(32b)$$

\noindent and then subjecting the ${\it totality}$ of the gauge theory
to the
nonunitary transform $U\times U^{\dagger}$. The method then yields:
the isounit
$I\rightarrow {\hat I} = U\times I\times U^{\dagger}$;
the isonumbers
$a\rightarrow {\hat a} = U\times a\times U^{\dagger} =
a\times (U\times U^{\dagger})
= a\times {\hat I}, a = n, c$;
the isoproduct with the correct expression and Hermiticity
of the isotopic element, $A\times B\rightarrow
U\times (A\times B)\times U^{\dagger} =
(U\times A\times U^{\dagger})\times (U\times
U^{\dagger})^{-1}\times (U\times B\times
U^{\dagger}) = {\hat A}\times {\hat T}\times {\hat B} =
{\hat A}{\hat {\times}}{\hat B}$;
the correct form of the isohilbert product on ${\hat C}$,
$<\phi|\times|\psi>\rightarrow
U\times <\phi|\times|\psi>\times U^{\dagger} =
(<\psi|\times U^{\dagger})\times (U\times
U^{\dagger})^{-1}\times (U\times |\psi>)\times (U\times U^{\dagger})
= <{\hat {\phi}}|\times {\hat T}\times |\hat {\psi}>\times {\hat I}$;
the correct Lie-Santilli isoalgebra $A\times B -
B\times A\rightarrow {\hat A}{\hat {\times}}{\hat B} -
{\hat B}{\hat {\times}}{\hat A}$;
the correct isogroup
$U\times (e^X)\times U^{\dagger} = (e^{X\times {\hat T}})\times {\hat
I}$,
the
isopoincare' symmetry ${\cal P}\rightarrow {\hat {\cal P}}$,
and the isogauge group $G\rightarrow {\hat G}$.

It is then easy to
verify that the emerging IGU is indeed invariant under all possible
additional
nonunitary transforms $W\times W^{\dagger} = {\hat I}$ provided that,
for evident reasons of consistency,  they
are written in their identical isounitary form,
$W = {\hat W}\times {\hat T}^{1/2}, W\times W^{\dagger} =
{\hat W}{\hat {\times}}{\hat W}^{\dagger} =
{\hat W}^{\dagger}{\hat {\times}}{\hat W}
= {\hat I}$.
In fact, we have the invariance of the isounit
${\hat I}\rightarrow {\hat I}' =
{\hat W}{\hat {\times}}{\hat I}{\hat {\times}}{\hat W}^{\dagger} = {\hat
I}$,
the invariance of the isoproduct
${\hat A}{\hat {\times}}{\hat B}\rightarrow
{\hat W}{\hat {\times}}({\hat A}{\hat {\times}}{\hat B})
{\hat {\times}}{\hat W}^{\dagger} = {\hat A}'{\hat {\times}}{\hat B}'$,
etc.
Note that the isounit is ${\it numerically}$ preserved, as it is the
case for
the conventional unit I under unitary transform, and that the selection
of
a nonunitary transform $W\times W^{\dagger} = {\hat I}'$ with
value different from $ {\hat I}$
evidently implies the transition to a different gravitational model.

Note that the lack of implementation of the above nonunitary-isounitary
lifting to only ${\it one}$ aspect of the original gauge theory (e.g.,
the
preservation of the old numbers or of the old differential calculus)
implies
the loss of the invariance of the theory$^{(4f)}$. The assumption of the
negative-definite isounit ${\hat I}^d = - (U\times U^{\dagger})$ then
yields
the isodual component of the IGU.

In closing, the most significant possibility we would like to convey is
that
{\it gravitation has always been present in unified gauge theories. It
did
creep in un-noticed because embedded where nobody looked for, in the
"unit"
of gauge theories}. In fact, the isogauge theory of Sect. 2 coincides
with the conventional theory at the abstract level to such an extent
that
we could have presented the proposed IGU
with exactly the same symbols of the conventional
gauge theories without the "hats", and merely subjecting the same
symbols to
a more general realization.

Also, the isounit representing gravitation
as per rule (32) verifies all the properties
of the conventional unit I of gauge theories, ${\hat I}^{\hat n} = {\hat
I},
{\hat I}^{\hat {1/2}} = {\hat I}, d{\hat I}/dt =
{\hat I}{\hat {\times}}{\hat H} - {\hat H}{\hat {\times}}{\hat I} =
{\hat H} - {\hat H} = 0$, etc.
The "hidden" character of gravitation in conventional gauge theories is
then confirmed by the isoexpectation value$^{(4f)}$
of the isounit which recovers
the conventional unit I of gauge theories,
${\hat <}{\hat I}{\hat >} =
<\hat {\psi}|\times {\hat T}\times {\hat I}\times {\hat T}\times |{\hat
{\psi}}>/
<\hat {\psi}|\times {\hat T}\times |\hat {\psi}> = I.$

It then follows that the proposed IGU constitutes an explicit
and concrete realization
of the theory of "hidden variables"$^{(13a)}$ $\lambda =
 T(x) = g(x)/\eta, \hat H\hat {\times}|\hat {\psi}> =
\hat H\times \lambda \times |{\hat \psi}> =
E_{\lambda}\times |\hat {\psi}>$,
and the theory is correctly reconstructed
with respect to the new unit
$ {\hat I} = {\lambda}^{-1}$, in which von Neumann's Theorem$^{(13b)}$
and Bell
s inequalities$^{(13c)}$ do not apply, evidently because of the
nonunitary
character of the theory (see Vol. II of Refs. [6g] for details).

In conclusion, as indicated
beginning with the title of the recent memoir$^{(4f)}$,
the proposed inclusion of gravitation in unified gauge theories is
essentially along the teaching of Einstein, Podolsky and Rosen$^{(14)}$
on the "lack of completion" of quantum mechanics,
only applied to gauge theories.

{\bf Acknowledgments}. I have no word to express my deepest
appreciation to Larry P. Horwitz of Tel Aviv University, for invaluable
critical comments and encouragements without which
this paper will not have seen the light of day. It is also a pleasant
duty to
express my deepest appreciation to: A. van der Merwe, Editor of
{\it Foundations of Physics}; P. Vetro, Editor of
{\it Rendiconti Circolo Matematico Palermo}; G. Langouche and H. de
Waard,
Editors of
{\it Hyperfine Interactions},  V. A. Gribkov, Editor of {\it Journal of
Moscow
physical Society}, and B. Brosowski, Editor
{\it Mathematical Methods in Applied Sciences}, for very accurate
Editorial
controls of the related publications$^{(4f,4g,7c,6e,10c)}$ which have a
fundamental
character for the results of this note. Thanks are finally due to J. T.
Goldman
of Los Alamos National Laboratories for the courtesy
of bringing to my attention ref. [12c].

 \end{document}